\documentclass[preprint,groupedaddress,endfloats*,amsmath,showpacs]{revtex4}
\usepackage[dvips]{graphicx,color}

\begin{document}

%\preprint{}

\title{Different pathways in mechanical unfolding/folding cycle of a single
semiflexible polymer}


\author{Natsuhiko Yoshinaga and Kenichi Yoshikawa}
%\thanks{}
\affiliation{Department of Physics, Graduate School of Sciences, Kyoto
University, Kyoto 606-8502, Japan}

\author{Takao Ohta}
\affiliation{Yukawa Institute for Theoretical Physics, Kyoto University, Kyoto 606-8502, Japan}
%\affiliation{}

\date{\today}

\begin{abstract}
Kinetics of conformational change of a semiflexible polymer under
 mechanical external field were investigated with Langevin dynamics
 simulations.
It is found that a semiflexible polymer exhibits large hysteresis in
 mechanical folding/unfolding cycle even with a slow operation, whereas
 in a flexible polymer, the hysteresis almost
 disappears at a sufficiently slow operation.
This suggests that the essential features of the structural transition of a semiflexible polymer
 should be interpreted at least on a two-dimensional phase space.
The appearance of such large hysteresis is discussed in relation to
 different pathways in the loading and unloading processes. 
By using a minimal two-variable model, the hysteresis loop is described in
 terms of different pathways on the transition between two stable states.
\end{abstract}

%macromolecules and polymer molecules
%Mechanical properties} 
%Phase transitions: general studies

%\keywords{}

\maketitle


\section{Introduction}

   Recent developments in experimental techniques regarding
   single molecule observation and manipulation have enabled us to
   directly investigate the statistical properties of individual small systems.\cite{svoboda:1994,bustamante:2003} 
These experiments have provided novel insights into the structure and function
of biological macromolecules.
These studies have revealed significant effects of fluctuation on the
properties of single molecules.

Most biopolymers, including DNA and many proteins, have finite
stiffness and are therefore classified as semiflexible polymers.
It is expected that the stiffness plays an important role in
the structural stability and function of these biomacromolecules.
Recent experiments and simulations have clarified that a
single semiflexible polymer exhibits first order phase transition between
elongated coil and folded compact states with a decrease in 
the solvent quality.\cite{Takahashi:1997} 
A single semiflexible polymer folds into various kinds of ordered structures
depending on its stiffness and temperature, such as a
toroid or a rod.\cite{noguchi:1996} 
This implies that the density of monomers is not sufficient to
characterize ordered structures, but the orientational order plays an essential role in the folding
transition of a semiflexible chain, i.e., the density-order coupling should be crucial.

%\textcolor[named]{Red}{
Among the studies on single molecule manipulation, experiments
on mechanical unfolding with laser tweezers or atomic force microscopy have been
actively performed over the past decade. \cite{rief:1997,baumann:2000,murayama:2001}
In these experiments, a non-trivial force response, which is called
a {\it{sawtooth-like}} pattern or a {\it{stick-and-release}} pattern, has
been observed either in a protein or in a DNA molecule.
Some authors claim that such patterns are associated with repeated domain
structures in a molecule.\cite{rief:1998,wada:2002}
In contrast to this, it is found in
\cite{baumann:2000,murayama:2001} that a DNA molecule having no repeated
sequence exhibits a {\it{stick-and-release}} pattern at the collapsing
conditions where the concentration of multivalent cations is high,
whereas the mechanical response of a worm-like chain is observed when
a DNA molecule is in the coil
conditions at the low concentration of multivalent cations.
Furthermore, it has been argued theoretically that a toroidal structure
of a folded semiflexible polymer without any repeated domains leads to
a {\it{stick-and-release}} pattern.\cite{kulic:2004,marenduzzo:2004}
%}

A system under loading and unloading processes is clearly in a
nonequilibrium state unless the rate of change is infinitesimally slow, and therefore the kinetics play
essential roles.
Recently, it has been revealed that a difference in the equilibrium free energy can be calculated from
kinetic processes under a nonequilibrium
state\cite{jarzynski:1997,crooks:1998} and the theory has been applied
to biopolymer stretching.\cite{hummer:2001} 
Despite such an analysis of nonequilibrium effects, there have been few
studies on the large hysteresis loop observed in the loading and
unloading cycle of biomolecules.

Hysteresis is a measure of how the system is away
from thermal equilibrium.
At the length scale of micron or submicron,
thermalization is so fast that equilibrium states are easily
attained.
However, to perform work against an external force, as seen with
molecular motors in biological systems,
the system should remain in a nonequilibrium state.
Therefore, it is important to clarify the origin of hysteresis and, especially, its robustness.

Theoretical and numerical studies are needed to gain general insight into
not only 
the nonequilibrium characteristics of a single polymer chain, but also
the working mechanism of a macromolecule in biological systems.
In the present article, we consider flexible and semiflexible polymers under
strain by molecular dynamics simulations.
We investigate the force response and focus on the close connection between hysteresis and the structural transition of a
single polymer chain.

A coarse-grained approach is generally quite useful to understand the
essence of the phase transitions. We shall introduce a Ginzburg-Landau type
free energy and the kinetic equations for the present problem.
As mentioned above, however, the density of monomers is not sufficient to
distinguish the different conformation in a semiflexible chain. Therefore, we
introduce another kinetic variable corresponding to the orientational order.
We do not intend to derive the free energy starting from the chain model but
show that the coarse-grained free energy having two metastable states indeed
describes properly the hysteresis in the force-response relation. To our
knowledge, this kind of approach has not been available in the kinetics
of a single chain.

In the next section, we present our chain model and describe the method of
simulations. In section III, the results of simulations are given. A
phenomenological scaling argument is developed in section IV. The
two-variable model and its numerical studies are shown in section V. A
summary is given in section VI.

\section{Simulations}   
To examine the folding and unfolding kinetics, we carried out Langevin dynamics
simulations of a bead-spring model using the following potentials
\begin{eqnarray} 
V_{\mathrm{beads}} &=& \frac{k _a}{2} \sum _{i} ({| \mathbf{r} _{i+1} -  \mathbf{r} _{i}|} - a)^2\label{V_beads}\\
V_{\mathrm{bend}} &=& \frac{\kappa}{2} \sum _{i}  (1 - \cos \theta _{i} )^2\\
V_{\mathrm{LJ}} &=& 4 \epsilon \sum _{i,j} ((\frac{a}{|\mathbf{r} _i -
 \mathbf{r} _j|})^{12} - (\frac{a}{|\mathbf{r} _i - \mathbf{r}
 _j|})^{6}),
\label{V_lj}
\end{eqnarray} 
where $V = V_{\mathrm{beads}} + V_{\mathrm{bend}} + V_{\mathrm{LJ}}$ and
$\mathbf{r} _{i}$ is the coordinate of the $i$th monomer and $\theta
_{i}$ is the angle between adjacent bond vectors.
The monomer size $a$ and $k_{B} T$ are chosen as the unit length and energy,
respectively, where $k_B$ is the Boltzmann
constant and $T$ is the absolute temperature. 
Monomer-monomer interaction is included by the Lennard-Jones potential
controlled by $\epsilon$.
A similar bead-spring model (\ref{V_beads}) - (\ref{V_lj}) has been used to study the properties of
semiflexible chains by setting a large value of $k_a$.\cite{sakaue:2002}
We adopt the parameters to be $k_a = 400$ and $\epsilon = 1.0$.
The bending elasticity is chosen to be $\kappa = 100$ for a semiflexible
polymer and $\kappa = 0$ for a flexible polymer.
The persistence length $l_p$
is a convenient measure to characterize the stiffness of a polymer chain.
In the present model, $l_p \sim 13.5 a$ for a semiflexible polymer.
Because of discritization of a chain with beads,
the minimum of the persistence length is $a$, and therefore $l_p \sim a$
for a flexible polymer.
We consider a homopolymer with a polymerization index $N = 200$.

A remark is now in order.
Because a semiflexible chain is locally inextensible, the partition
function should be written as $Z = \int  {\mathcal{D}} {\bf r}_i  \Pi \delta (| \mathbf{r} _{i+1} -  \mathbf{r} _{i}| - a) e^{-(V_{\mathrm{bend}} + V_{\mathrm{LJ}})}$.
As an analytically or numerically tractable model, several approximations have
been employed and examined.
For example, the local constraint was replaced by a global constraint in
refs.\cite{winkler:2003,ha:1995} to study analytically the equilibrium properties of a semiflexible
polymer.
Instead of such approximations, we adapt the harmonic potential
(\ref{V_beads}) between a pair of the neighboring beads.
A continuum version of the model (\ref{V_beads}) was introduced to
evaluate the end-to-end distance\cite{saito:1967} which shows clearly
that the end-to-end distance is insensitive to $k_a$ for large values of
$k_a$.

%However we connect the neighboring beads by the harmonic potential
%(\ref{V_beads}) instead of using the constrain by the delta function.
%This leads to elongation and contraction of bond length.
%In general, such elongation (or contraction) and bending are coupled,
%but for large $k_a$ they are approximately decoupled.\cite{saito:1967,soda:1973}
%Therefore we may assume that bond length is constant.
%In addition, the relaxation time for the fluctuation of elongation and
%contraction is much faster than other characteristic time scales of our
%interest.
%Thus we can justify the harminic potential (\ref{V_beads}) to connect
%neighboring beads.

The equation of motion can be written as,
\begin{equation}
m \frac{{\mathrm{d}}^2 {\mathbf{r}} _i}{{\mathrm{d}} t^2} 
= - \gamma \frac{{\mathrm{d}} {\mathbf{r}} _i}{{\mathrm{d}} t}
- \frac{\partial V }{\partial {\mathbf{r}} _i}
+ {\boldsymbol{\xi}} _i,
\label{simulation}
\end{equation}
where $m$ and $\gamma $ are the mass and
friction constant of a monomeric unit, respectively.
The constant $\tau = \gamma a^2/k_{B}T$ is chosen to be the
unit for the time scale.
We set the time step as $0.01 \tau$, and use $m=1.0$ and $\gamma =1.0$.
Gaussian white noise ${\boldsymbol{\xi _{i}}}$ satisfies a 
fluctuation-dissipation relation,
\begin{equation}
 <{\boldsymbol{\xi}} _i (t) \cdot {\boldsymbol{\xi}} _j (t')> = 6 \gamma
  k_{B}T \delta _{ij} \delta (t-t'),
\end{equation}
where $<\cdots>$ indicates an ensemble average.

The folding and unfolding states of a polymer are characterized by the average local
monomer density $\rho$,
which is defined by,
\begin{equation}
\rho = \frac{1}{N} \sum _{i,j} H ( r_c ^2 - | {\mathbf{r}}_i - {\mathbf{r}}_j |^2),
\end{equation}
where $H (x)$ is the Heaviside function.
In this work, we set $r_c = 3.0$.

Several comments on the values of the parameters chosen above are in
order.
The attractive strength $\epsilon$ in the Lennard-Jones potential
(\ref{V_lj}) is chosen so that ordered structures such as toroids and
rods are formed for polymerization index $N=200$ and $l_{p} \sim
13.5a$.\cite{noguchi:1996}
Among such ordered structures, a toroidal structure is the most
energetically favorable in the present set of parameters.
The spring constant $k_{a}$ in (\ref{V_beads}) is set to be a
sufficiently large value to realize the local inextensibility of a chain.
We set $m = \gamma = 1.0$ in (\ref{simulation}) to optimize the
numerical simulations within accessible simulation time avoiding any
numerical instabilities.
With this condition, the inertia term does not cause any artifacts because
the relaxation time of the momentum of a monomer is
sufficiently faster compared to that of conformational change.

We fix one end of a polymer chain and move the other end at constant
velocity $\dot{z}$.
The end-to-end distance $z$ is one of the measurable quantities. 
The force $f$ applied to the $N$th monomer is monitored during the operation.
Note that force is averaged over time to avoid its large fluctuation.
The inverse velocity $\lambda = \dot{z}^{-1}$ is often used as a measure of the speed of
external operations.
It is convenient to define $\lambda _{\mathrm{fold}}$, which corresponds
to the time scale of the folding transition of a semiflexible polymer in
bulk.
It can be written as $\lambda _{\mathrm{fold}} = <\tau _{\mathrm{fold}} /(R
_{\mathrm{coil}} - R _{\mathrm{fold}} )>$, where $\tau _{\mathrm{fold}}$ is
the characteristic time for a folding transition, and $R
_{\mathrm{coil}}$ and $R _{\mathrm{fold}}$ are the end-to-end distance
of the coiled and folded states in a semiflexible polymer, respectively.

Here we mention the time and length scales in our simulations.
The advantage of the bead-spring model
(\ref{simulation}) with (\ref{V_beads}-\ref{V_lj}) is that we can
simulate on a realistic time scale. 
The folding transition time is on the order of $ 10^5 - 10^6$ steps, which corresponds to about 0.1 - 1.0 sec in the experiment with DNA.\cite{yoshikawa:1996b}
From this correspondence, our loading and unloading speed ($ 10^7 -
10^8$ steps in the simulations) is considered
to correspond to 10 - 100 sec, which is a
experimentally accessible value.
We can also discuss the length scale.
The persistence length of DNA is about 50 nm, which corresponds to about
10$a$ in our simulation.
Therefore, we can estimate the size of a toroidal state
of a semiflexible polymer ($\sim 10a$ in the simulation) to be about 50 nm, which is in good agreement
with experiments on DNA. 
Most simulations suffer from a large value of the force response.\cite{li:2003}
However, the force response in our simulations ($2a-5
a/k_{B}T$) corresponds to $3-7 {\mathrm{pN}}$, which is consistent with the experimental value. \cite{murayama:2001}

\section{Results}

\begin{figure}
\includegraphics{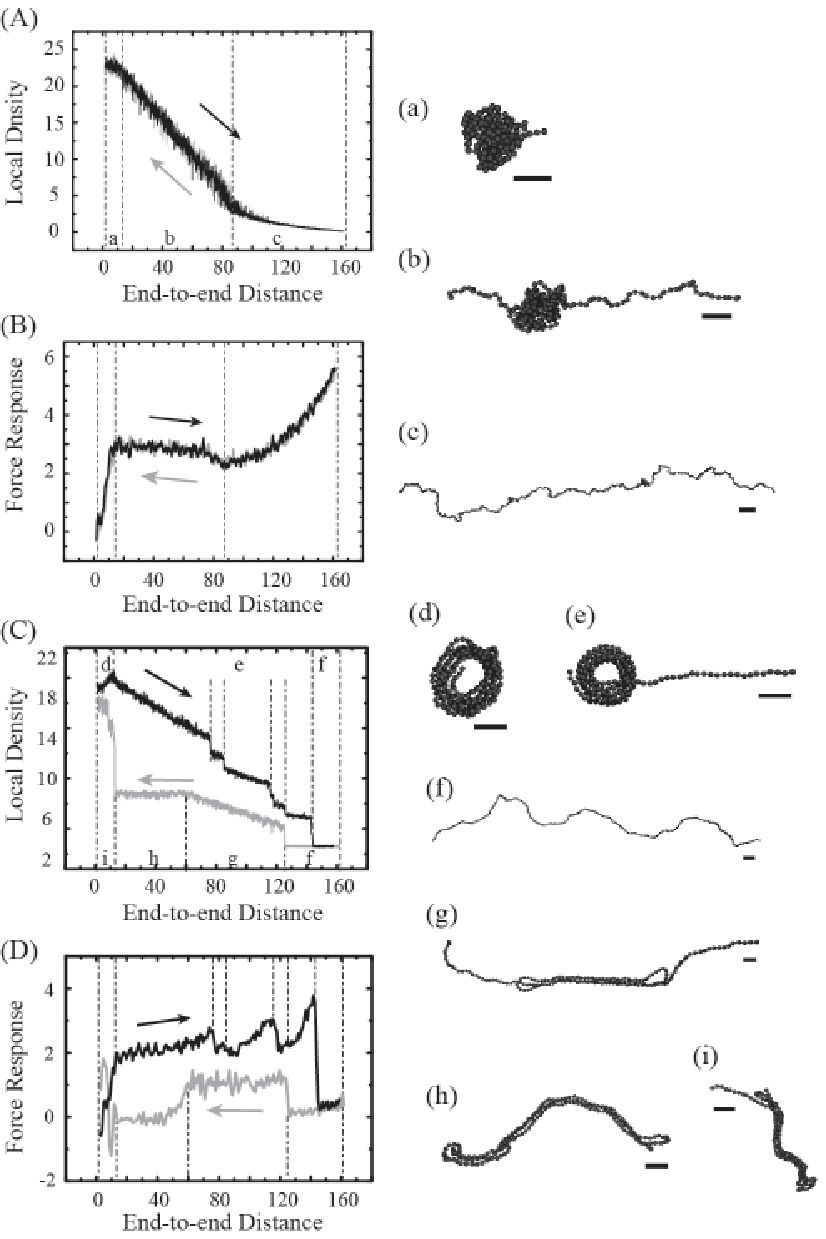}
\caption{Local monomer density and force
 response under strain in flexible (A, B) and semiflexible (C, D) polymers. The black and gray
 lines show loading and unloading processes respectively.
Typical snapshots of chain conformation are also shown. 
The scale bar corresponds to $5a$.
\label{fig:flexible}
}
\end{figure}

Figure \ref{fig:flexible} shows the local monomer density and the force response of a flexible polymer and a semiflexible polymer, where the former
takes a spherical globule morphology and the latter takes a toroidal morphology in
their compact states.
We have chosen the inverse velocity as $\lambda = 10 ^4 \tau / a$ in Fig.1.
It can be seen that the pathway of structural transition in the loading
of a semiflexible polymer is clearly different from that in unloading.

The force response in a flexible polymer increases at $z < 20$,
maintains a constant value at $20 < z < 80$, and then increases at $z > 80$.
Correspondingly, a globular flexible polymer changes to an ellipsoidal
shape and then to a phase-segregated state consisting of 
a coil and a globule, followed by a coil state.
This behavior is observed in both loading and unloading, and
thus a flexible polymer does not exhibit hysteresis. 
The force response observed in Fig.\ref{fig:flexible}B(b) is called a {\it{force plateau}}, and was
predicted by Halperin et al.\cite{halperin:1991,wittkop:1996} 

In the loading of a semiflexible polymer, a toroidal semiflexible polymer shows phase segregation of a coil and a toroid part.
In this state, the monomer density decreases in a stepwise manner, and
    correspondingly the force response exhibits a {\it{stick-and-release pattern}}, which has been observed in experiments with DNA.\cite{baumann:2000,murayama:2001}
At $z \sim 140$, the polymer undergoes a transition to a coiled state.
In the unloading process, on the other hand, the phase-separated state
    of a rod and a coil appears at $z \sim 125$, and the polymer
    becomes a single phase consisting of a rod.
At the final stage, a rod becomes more folded or makes a transition to a toroid. 
The force response in a rod-coil phase-segregated state exhibits a
{\it{force plateau}} and is largely different from that in a toroid-coil state.

\begin{figure}
\includegraphics{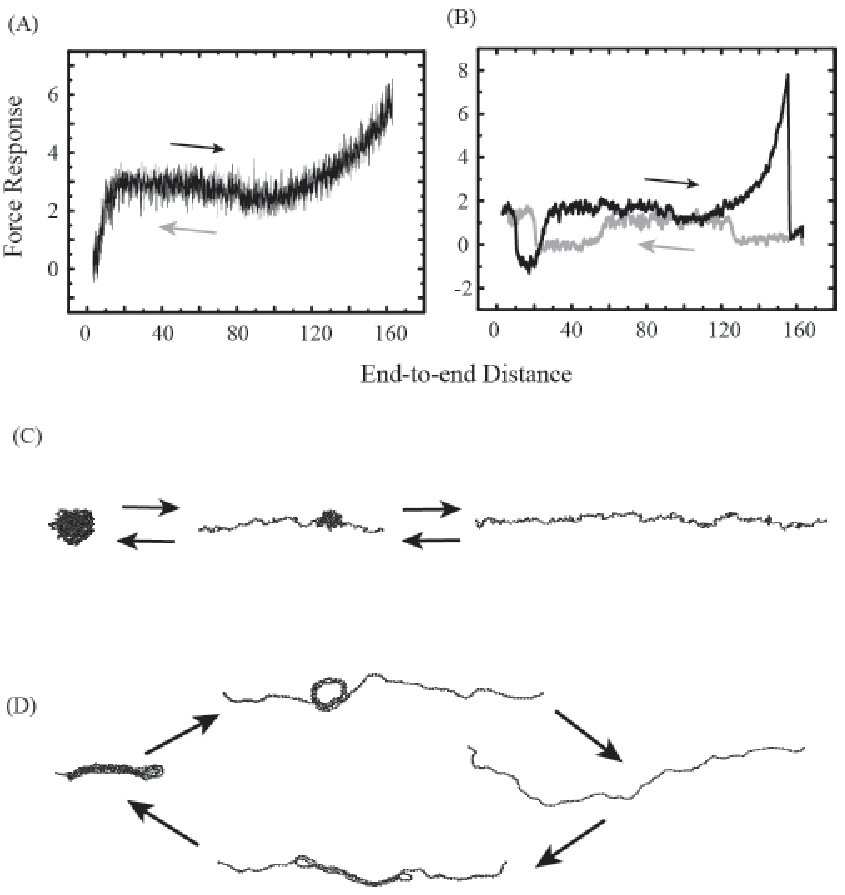}
\caption{Force response in flexible (A) and semiflexible
 (B) polymers. While the transition pathways in loading and unloading
 processes are the same for a flexible polymer (C), for a semiflexible polymer
 the pathway in loading is largely different from that in
 unloading (D).
\label{fig:stiff}
}
\end{figure}

\begin{figure}
\includegraphics{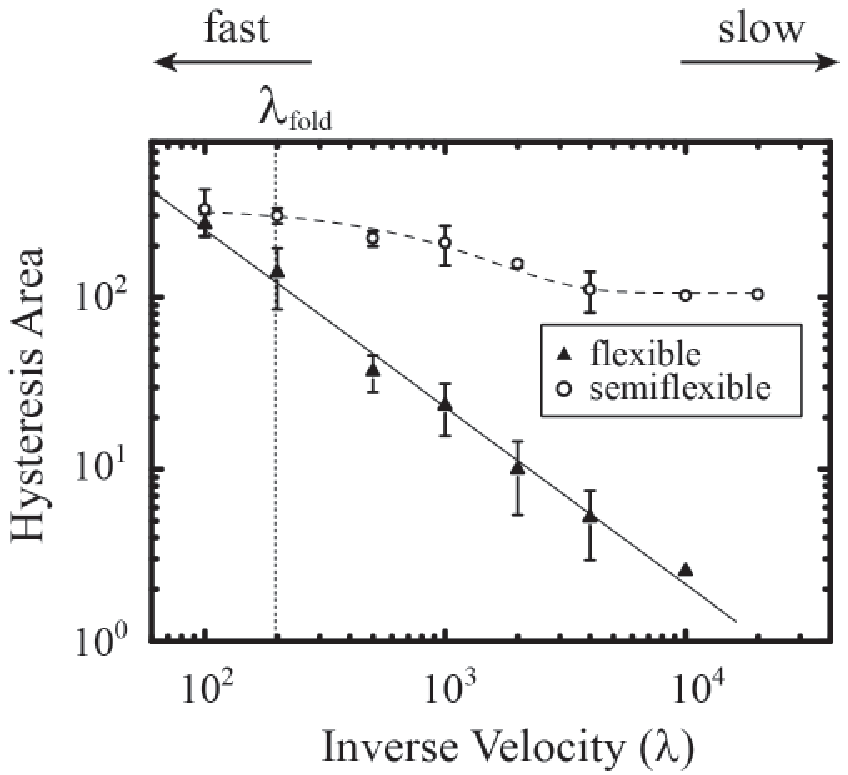}
\caption{Inverse velocity dependence on hysteresis in flexible (closed triangles) and semiflexible (open circles) polymers in our
 simulations.
The solid and broken lines are guides to eyes.
\label{fig:hysteresis_simu}
}
\end{figure}

Next, we consider the
cycle in which the initial and final states are coiled states.
As shown in Fig.\ref{fig:stiff}, a semiflexible polymer exhibits a large hysteresis
loop, in contrast to a flexible polymer.
To see the difference more quantitatively, we calculate the hysteresis area
at various operating speeds.
Figure \ref{fig:hysteresis_simu} shows the hysteresis area of the force response, i.e. the difference
in dissipative work between loading and unloading processes.
In a flexible polymer, hysteresis rapidly decreases as the operating
speed decreases.
In contrast, hysteresis in a semiflexible
polymer decreases rather slowly.
Large hysteresis remains even at $\lambda \sim 10^{4} \tau / a$, although
a semiflexible polymer in bulk makes a transition to a folded state on the time
scale $\lambda _{\mathrm{fold}} \sim 2 \times 10^2 \tau / a$.
Therefore, we conclude that a semiflexible polymer maintains large hysteresis
even at an extremely slow operation speed.

The appearance of notable hysteresis in a semiflexible polymer is associated with
the significant difference in the pathways between the loading and
unloading processes.
In unloading, a rod-like structure appears frequently, whereas
in loading, a toroidal loop structure tends to be generated.
As shown in Figs.\ref{fig:flexible} and \ref{fig:stiff}, the force response strongly depends on the
structure of the folded part.
This structural difference is the origin of the large hysteresis in a
semiflexible polymer.
In contrast, there is almost no difference in the transition pathways of
a flexible polymer between the loading
and unloading processes (Fig.\ref{fig:stiff}C).

Obviously, the hysteresis should disappear at the infinitesimally slow operation
where the transition between a toroid and a rod is achieved.
However in a practical sense, it is difficult to observe the transition
because of a large free energy barrier between these states.

\section{Scaling analysis}

In this section, we will discuss physical meanings of the hysteresis
appearing in the cycle.
Energy balance can be written with the work ${\mathrm{d}}W$ per unit
time by the external perturbation as,
\begin{equation}
 {\mathrm{d}}W = {\mathrm{d}} F + {\mathrm{d}} Q,
\end{equation}
where the first and second terms in the r.h.s. are free energy change
and dissipative heat, respectively.
The hysteresis area $A$ in the present system corresponds to the
difference between total works in loading and unloading process, $A =
|W_{\mathrm{load}}| - |W_{\mathrm{unload}}|$.
The dissipative heat is equivalent to the entropy production of solvent
molecules,
\begin{equation}
 {\mathrm{d}} Q = {\mathrm{d}} S.
\end{equation}
In the estimation of the entropy production, only
the Stokes drag is considered because the hydrodynamic interaction is
neglected in the present work.

We consider a polymer which consists of coiled
and collapsed parts.
The coiled part has $n$ unfolded monomers, and is $l$ in size in the
direction parallel to the external force and $R$ in the
perpendicular direction.
The collapsed part is either globule, toroid or rod state.
The main contribution to the entropy production comes from the coiled
part and is
\begin{equation}
 \frac{{\mathrm{d}} S}{{\mathrm{d}} t}
\sim \eta n a (\frac{{\mathrm{d}} R}{{\mathrm{d}} t})^2,
\end{equation}
where $\eta$ is the viscosity.\cite{halperin:2000}
The monomer size $a$ is hereafter set to be unity.
Assuming that $n$ monomers behave as a Gaussian statistics in the perpendicular direction to the
external force, i.e. $R \sim n^{1/2}$, we obtain 
\begin{equation}
 \frac{{\mathrm{d}} S}{{\mathrm{d}} t}
\sim \eta  \dot{z}^2, \label{dS_dt}
\end{equation}

In a flexible polymer, we can evaluate the free energy change to be,
\begin{equation}
  \frac{{\mathrm{d}} F}{{\mathrm{d}} t}
\sim (\epsilon - \frac{l^2}{n^2 }) \dot{z}.
\end{equation}
The free energy change is reversible between loading and unloading
process, and therefore it does not contribute to the hysteresis.
As a result, the hysteresis in a flexible polymer can be expressed as
\begin{equation}
 A 
\sim 
\eta  \dot{z} N \sim \frac{\tau _{\mathrm{R}} \dot{z}}{N},
\label{hysteresis_theory}
\end{equation}
where $\tau_{\mathrm{R}}$ is the Rouse relaxation time of a polymer.

In a semiflexible polymer, the collapsed parts have different structures
between loading and unloading process, and therefore the free energy
change contributes to the hysteresis.
The free energy in a loading and unloading
process can be written as the summation of three contributions; the surface,
bending and volume free energy.
\begin{eqnarray}
 F_{\mathrm{load}} 
&\sim& \mu d_1 r_1 + l_p \frac{N-n}{r_1 ^2} - \epsilon  (N -n),\\
 F_{\mathrm{unload}} 
&\sim& \mu d_2 r_2 + l_p \frac{d_2 ^3}{d_2 ^2} - \epsilon (N -n),
\end{eqnarray}
where $\mu$ is the surface free energy density.
A toroid is regarded as a ring with the radius $r_1$ and the width
$d_1$, whereas a rod is characterized by the length $r_2$ and the width $d_2$.
From the conservation of the volume of a collapsed part, the relations $d_1 \sim
((N-n)/r_1)^{1/2}$ and $d_2 \sim ((N-n)/r_2)^{1/2}$ have to be
satisfied.
At a given $n$, we may assume that the collapsed part rapidly thermalizes,
and therefore the surface and bending terms should be balanced.
As a result, the free energy in a loading and unloading process is given by
\begin{eqnarray}
 F_{\mathrm{load}} 
&\sim& l_p ^{1/5} (N-n)^{3/5} \mu ^{4/5} - \epsilon  (N -n),
\label{F_semi_toroid}\\
 F_{\mathrm{unload}} 
&\sim& l_p ^{1/2} (N-n)^{1/2} \mu ^{1/2} - \epsilon (N -n),
\label{F_semi_rod}
\end{eqnarray}

The entropy production in a semiflexible polymer also arises from
the coiled part.
%When the persistence length is much larger than the size of a thermal
%blob, {\it i.e.} $l_p \gg \xi _c$, 
Approximating that the coiled part consists of cylinders with length
$l_p$, we estimate the entropy production as\cite{doi:1986}  
\begin{equation}
 \frac{{\mathrm{d}} S}{{\mathrm{d}} t}
\sim \frac{n}{l_p} \frac{\eta l_p}{\ln l_p} (\frac{{\mathrm{d}} R}{{\mathrm{d}} t})^2,
\end{equation}
which is valid $l_p \gg 1$.
The bending energy at the length scale $n$ is of the order of the thermal
energy $k_B T$ so that $l_p (R/n)^2 \sim 1$.
As a result, we obtain
\begin{equation}
 \frac{{\mathrm{d}} S}{{\mathrm{d}} t}
\sim \frac{\eta n \dot{z}^2}{l_p \ln l_p},
\label{dS_dt_semi} 
\end{equation}
From (\ref{F_semi_toroid}), (\ref{F_semi_rod}) and (\ref{dS_dt_semi}),
the hysteresis area in a semiflexible polymer can be estimated as
\begin{eqnarray}
 A &\sim&
\Delta F + \frac{ \eta N^2 \dot{z}}{l_p \ln l_p},
\label{hysteresis_semi_theory}\\
\Delta F &=& 
l_p ^{1/5} N^{3/5} \mu ^{4/5} 
- l_p ^{1/2} N^{1/2} \mu ^{1/2}.
\end{eqnarray}
When $\dot{z}$ is small, i.e., the operating speed is small (but still
large such that the
probability of the transition between a rod and a toroid
is almost zero), the difference between the free energy change in loading and unloading
process is larger than the entropy production.
Therefore, the hysteresis is almost insensitive to the operating speed.
The above results are qualitatively consistent with those of
our simulations (Fig.\ref{fig:hysteresis_simu}).
As shown in (\ref{hysteresis_theory}), the hysteresis area in a flexible
polymer decreases linearly as $\lambda = \dot{z} ^{-1}$ increases.
On the other hand, (\ref{hysteresis_semi_theory}) indicates that the
hysteresis in a semiflexible polymer is almost insensitive to $\lambda $
when $\lambda $ is large.

In this estimation, we neglect the contribution of the relaxation from a
rod to a toroid as well as from a rod to a more folded rod. However, this
process occurs only at small values of $z$
and hence does not contribute to the hysteresis when $N$ is large.

\section{Two-variable model}

Hysteresis is a general property of a first order phase transition.\cite{landau:1958} 
However, the hysteresis described in the preceding section is different
from that in the ordinary case. 
In a semiflexible polymer, the kinetic pathway in the loading process
is different from that in the unloading process. 
This is because different structures appear during the transition
processes. 
Therefore, one has to consider two metastable states in the kinetics of
the folding and unfolding transitions of a semiflexible polymer chain. 
That is, a transition from a folded state to an unfolded state occurs
via one metastable state whereas the system traverses another metastable
state in the reverse process. 
In order to represent this kind of phase transition, one has to
introduce at least two variables in the kinetic equations.

In this section, we consider a model system that provides the above
mentioned characteristic features. 
We do not intend to develop a quantitative theory specific to a
semiflexible polymer chain, but rather study the hysteretic property from a
general point of view based on a simple Ginzburg-Landau type approach.

We mention a recent theoretical study based on
the similar idea.
Bartolo et al. have investigated dynamic response of adhesion complexes
by using the multidimensional energy landscape.\cite{bartolo:2002} 
What they have found is that two alternative trajectories are possible
depending on the loading rate.
This is different from our concern that the system relaxes to the most
stable state via two different metastable states when the process is
reversed under the same loading rate.

We start with the free energy in terms of two order parameters $X$
and $Y$. 
\begin{equation}
F(X,Y) = -\frac{\mu_1}{2}X^2-\frac{\mu_2}{2}Y^2 - \alpha XY+\frac{1}{4}X^4+
	 \frac{1}{4}Y^4-h(t)X,
\label{modelfree}
\end{equation}
where $\mu_1$, $\mu_2$
	 and $\alpha $ are positive
	 coefficients and $h$ is a
	 control parameter depending on time. 
For a suitable set of parameters, the free energy has four local
	 minima, two of which correspond to folded and unfolded
	 states. 
Qualitatively, the variables $X$ and $Y$ correspond, respectively, to the density and the orientational order of a polymer with a suitably chosen origin of the scale.
The coupling between them is incorporated into the third term in the free energy (\ref{modelfree}) such that increasing the density increases the orientational order and vice versa.

The molecular dynamics simulations in the
preceding section were carried out by changing the end-to-end distance
at a constant speed.
That is, the end-to-end
	 distance was changed and the force exerted there was measured. 
To consider the corresponding situation, we regard the external parameter $h(t)$ as the end-to-end distance and assume that 
there is a one-to-one correspondence between the density and the force response.
Under these circumstances, we evaluate the hysteresis area in $X$ by changing the external parameter.
We fix one parameter $\mu _1$ to be $\mu _{1} = 100$ and
examine two situations $\mu _{1} \gg
	 \mu _{2} = 0.1$ and $\mu _{1} = \mu _{2} = 100$.
When $\mu _{2} = 0.1$, there are only two stable minima in the free
energy and hence the $Y$ variable is irrelevant and corresponds to an ordinary flexible polymer.
In contrast, it will be shown below that the case with $\mu _{2} = 100$ corresponds to a semiflexible polymer.

The kinetic equations for $X$ and $Y$ are assumed to be given,
	 respectively, by
\begin{eqnarray}
&& \frac{{\rm{d}} X}{{\rm{d}} t} = -L_{1}\frac{\partial F}{\partial
 X}+\xi _{1}
=L_{1} (- X^3 + \mu _{1}  X + \alpha Y + h(t))
  +\xi
 _{1} \label{modelx}\\
&& \frac{{\rm{d}} Y}{{\rm{d}} t} = -L_{2}\frac{\partial
 F}{\partial Y}+ \xi
  _{2}
=L_{2} (- Y^3 + \mu _{2} Y + \alpha X )+
 \xi
  _{2}\label{modely},
\end{eqnarray}
where $L_1$ and $L_2$ are the Onsager coefficients. 
The Gaussian random forces $\xi_1$ and $\xi_2$ satisfy the
fluctuation-dissipation relation,
\begin{equation}
<\xi _{i}(t) \xi_{j}(t')> = 2L_{i} k_{B}T \delta _{ij} \delta (t-t').
\end{equation} 
We carried out numerical simulations for a set of Langevin
equations with $L_1=L_2=1$ and $\alpha =5.0$. 
The external force is changed linearly in time;
\begin{equation}
h(t) = \pm vt+h_0,
\end{equation}
with $h_0$ an initial value.

\begin{figure}
\includegraphics{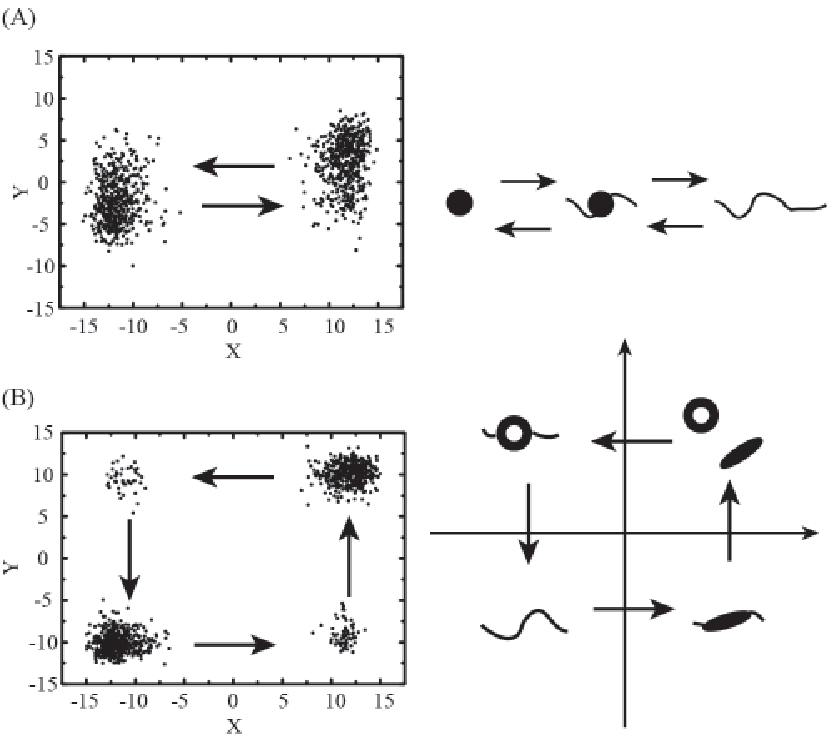}
\caption{Time trajectory of our model calculation in the XY plane for
 $\mu _2 = 0.1$ (A) and $\mu _2 = 100$ (B),
 and schematic representations of the correspondence to our
 simulation results in Fig. \ref{fig:stiff} (C) and (D).
\label{fig:model}
}
\end{figure}

\begin{figure}
\includegraphics{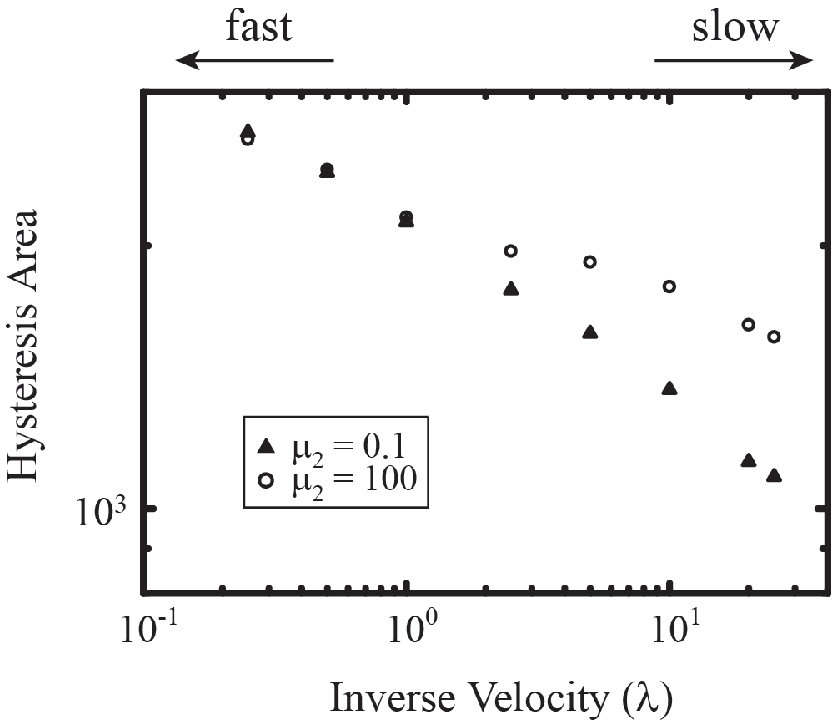}
\caption{Inverse velocity dependence on hysteresis at $\mu _2 = 0.1$
 (closed triangles) and $\mu _2 = 100$ (open circles) obtained
 from (\ref{modelx}) and (\ref{modely}).
\label{fig:hysteresis_model}
}
\end{figure}

Figure 4 displays the numerical results of (\ref{modelx}) and
(\ref{modely}).
When $\mu _2 = 0.1$, there are two stable equilibrium solutions at $(X,Y) \simeq (\pm 12,\pm 5)$.
With change in the external field $h(t)$, transitions occur from one state
to the other state.
This corresponds to the transition between a folded state and an
elongated state of a flexible polymer.
In fact, the hysteresis area decreases as the inverse velocity $\dot{z}^{-1}$ is
increased, as shown in Fig.\ref{fig:hysteresis_model}.
When $\mu _2 = 100$, two metastable states appear at $(X,Y)
\simeq (\pm 10, \mp 10)$ apart from the more stable states at $(X,Y) \simeq (\pm
10,\pm 10)$. 
The transition occurs through a different path via one of these
metastable states depending on an increase or a decrease
in the external field.
In this case, the hysteresis area does not decrease substantially for
a slow operation compared to the case with $\mu _2 = 0.1$, as shown in
Fig.\ref{fig:hysteresis_model}.
This behavior of hysteresis in Fig.\ref{fig:hysteresis_model} is indeed
consistent qualitatively with the results shown in Fig.\ref{fig:hysteresis_simu}
obtained by molecular dynamics simulations.

The scale of the horizontal and vertical axes in
Fig.\ref{fig:hysteresis_model} is much different from that in
Fig.\ref{fig:hysteresis_simu}. 
This arises from the uncertainty of the relation between the density and
the force response as mentioned above and from the difficulty of
achieving correspondence to the time scale of the folding of a polymer in the artificial model system (\ref{modelx}) and
(\ref{modely}).

\section{Summary}
In summary, we have studied the force-strain
relation of a bead-spring model of a polymer and found the large hysteresis in
the loading-unloading cycle for a semiflexible polymer.
What is different from the ordinary hysteresis is that a large
hysteresis is observed even with a slow
operation, which implies that there are different pathways of conformational change between folding and unfolding.
This is in contrast to a flexible polymer in which the hysteresis area
disappears almost entirely with a sufficiently slow operation. 
This means that the conformational change of a flexible polymer can be reduced essentially into
a one-variable problem.

In the present paper, we have considered two limiting cases of flexible
and semiflexible polymers.
It would be an interesting problem to investigate systematically the
change of the hysteresis by changing the bending elasticity $\kappa$.
However this is left for a future study.

Although most biopolymers such as proteins have much more complicated structures,
they often exhibit the characteristics of a semiflexible polymer.
For example, the secondary structures such as $\alpha $-helix and
$\beta$-sheet parts are rather stiff and, therefore, permit a locally ordered structure.
As a result, even if the initial and final states are the same, the transition paths
would exhibit a large hysteresis loop, as seen in a semiflexible polymer in the present paper.
Such kinetic properties are expected to play an important role in biomacromolecules.

\begin{acknowledgments}
This work is supported by the Grant-in-Aid for the 21st
Century COE "Center for Diversity and Universality in
Physics" from the Ministry of Education, Culture, Sports,
Science and Technology (MEXT) of Japan
%This work was supported in part by Grants-in-Aid for the 21st Century COE
% (Center for Diversity and Universality in Physics) 
and a fellowship
 from the JSPS fellows
 (No. 1142).
N. Y. thanks Prof. H. Schiessel, Prof. M. Sano, Dr. Y. Murayama and
 Mr. H. Wada for their helpful discussions.
\end{acknowledgments}

\bibliography{yoshinaga}

%\begin{thebibliography}{17}
%\expandafter\ifx\csname natexlab\endcsname\relax\def\natexlab#1{#1}\fi
%\expandafter\ifx\csname bibnamefont\endcsname\relax
%  \def\bibnamefont#1{#1}\fi
%\expandafter\ifx\csname bibfnamefont\endcsname\relax
%  \def\bibfnamefont#1{#1}\fi
%\expandafter\ifx\csname citenamefont\endcsname\relax
%  \def\citenamefont#1{#1}\fi
%\expandafter\ifx\csname url\endcsname\relax
%  \def\url#1{\texttt{#1}}\fi
%\expandafter\ifx\csname urlprefix\endcsname\relax\def\urlprefix{URL }\fi
%\providecommand{\bibinfo}[2]{#2}
%\providecommand{\eprint}[2][]{\url{#2}}
%\end{thebibliography}

\end{document}